\documentclass[twocolumn,showpacs,amsmath,amssymb,aps]{revtex4}
\usepackage{amsmath}
\usepackage{dcolumn}
\usepackage{bm}

\begin{document}

\title{Detectability of Lorentz-violating potentials in a unified model of fermions}

\author{Kimihide Nishimura}
\email{kmdns@tune.ocn.ne.jp}
\affiliation{Nihon Uniform, 1-4-21 Juso Motoimazato, Yodogawa-ku Osaka 532-0028 Japan}

\date{\today}

\begin{abstract}
The detectability of the fermion-potentials appearing in a unified model of fermions is discussed from the viewpoint of an effective field theory.
Although the fermion-potentials are effectively represented as terms similar to the $a$-coefficients in the theory of standard-model extension, their magnitudes are very large and their physical implications are different. 
A possibility is shown that the fermion-potentials are detectable by neither the deviations from conventional energy-momentum conservations, the neutrino-oscillations, the CPT-violation in neutral meson systems, nor the gravitational effects.
\end{abstract}

\pacs{12.10.-g, 11.30.Cp, 12.15Ff, 14.60Pq}
\maketitle

\section{Introduction}
A realistic model of spontaneous Lorentz violation (SLV) has been recently proposed \cite{KN}, which is expected to provide a grand unified theory beyond the standard model. It suggests that SLV may deeply participate in forming the structure of elementary particles and forces symbolized by the standard theory.

In this model, constant potential terms appear in dispersion relations for quasi fermions.
These terms, which we here call the fermion-potentials, or simply the $f$-potentials, can be expressed in an effective field theory by the terms similar to the $a$-type Lorentz-violating coefficients discussed in the standard-model extension (SME)  \cite{CK1,CK2}, and therefore appear to violate Lorentz covariance and CPT invariance. In contrast to the $a$-coefficients in SME, the fermion-potentials have very large values, nearly of the same order of magnitude as masses of  leptons and quarks. This characteristic seems to rise several problems.

The notion of SME itself was invented by Kosteleck\'{y} and his coworkers for examining the consequences from Lorentz violation in terms of an ordinary local quantum field theory. The physical effects from $a$-coefficients have been investigated in this context. 
In SME, the magnitudes of $a$-coefficients are severely constrained  by the experiments for Lorentz invariance \cite{KR}. 
From the resemblance between the $f$-potentials and the $a$-coefficients, it seems therefore that large $f$-potentials in the unified model of fermions would immediately contradict to experimental verifications of Lorentz invariance. Moreover, it can be shown that, if interpreted in the context of SME, large $f$-potentials would lead to an erroneous energy-momentum conservation law different from the conventional one in charged weak current processes.

On the first question, we can show at least in an effective free field theory that $f$-potentials in dispersion relations are removable by appropriate local phase transformations for Dirac spinors, as first argued in SME \cite{CK1}.  In this sense, the $f$-potentials do not cause any contradiction to experimental constraints for CPT and Lorentz invariance. 
In addition, even if interactions are taken into account, due to the nature of SLV, it seems difficult to imagine that the interactions would generate phenomena appreciably violating Lorentz invariance. 

However, if the unified model of fermions reproduces the standard model as an effective field theory, the second problem seems difficult to solve in the context of SME.

We then come to a conclusion that $f$-potentials in the unified theory of fermions can not be identified with, or interpreted as $a$-coefficients in SME.
This observation leads us to another notion, the standard-model redefinition (SMR), instead of SME.

There is a case in SME that $a$-coefficients are unconstrained by experimental measurements. This case occurs when they are eliminable by appropriate field redefinitions. This observation suggests that there are some Lorentz-violating models which really show no Lorentz-violating phenomenon.

An example is easily obtained if we perform for a Lorentz invariant theory some field redefinitions parameterized by tensor constants. Then the Lagrangian will appear to break Lorentz invariance. We may call the Lorentz violation of the model which can be transformed by appropriate field redefinitions into the Lorentz invariant one the ``trivial Lorentz violation" (TLV). 

The aforementioned SMR is obtained from the Lagrangian of the standard model by the local linear phase transformations for spinors. 
Namely, SMR is a TLV form of the standard model.

Though, in this case, the $f$-potentials seem not detectable at all by experimental observations, the detection of the existence of $f$-potentials has a great importance for verifying the reality of explanations given by the unified picture of fermions for the origin of various elementary particles and forces in nature.

In the previous paper \cite{KN}, it has been shown that, if $f$-potentials are effective in the matter-generating-era of the Universe, they can explain well the baryon asymmetry of the Universe under some natural assumptions. This paper further discusses of whether the existence of $f$-potentials are detectable in other phenomena.

The possibilities to detect $f$-potentials seem to remain in the gravitational effects and the flavor-mixings.
The former concerns an expectation for some connection between the $f$-potentials and the origin of dark matter and dark energy discussed in Cosmology \cite{Komatsu}. The latter may generate observable effects even in SMR, 
since the local phase transformations change particle-identifications according to the space-time points.
We examine these possibilities by considering the conservation of energy-momentum tensor and the neutrino oscillations in the quasi-fermion picture.

In addition, there are arguments in SME that $a$-coefficients obstruct the conservation of symmetric energy-momentum tensor, and that $a$-coefficients are not removable by the electromagnetic gauge transformation \cite{CK1}. We also examine corresponding questions in SMR.

\section{Effective field theory for quasi fermions}
According to the unified model of fermions \cite{KN}, the dispersion relations for quasi-leptons have the following forms 
\begin{equation}
\begin{array}{ll}
{\rm a\ quasi\ neutrino}&:p_0=|\bm{p}|\pm\mu_l,\\
{\rm a\ charged\ quasi\ lepton}&:p_0=\sqrt{\bm{p}^2+m_l^2}\mp\mu_l,\\
\end{array}
\label{LeptonDR}
\end{equation}
where the lower sign corresponds to a quasi-anti-lepton, and $l$ runs over $e$, $\mu$ and $\tau$.
Similarly, the dispersion relations for quasi-quarks are definable in the isotropic representation as
\begin{equation}
\begin{array}{ll}
{\rm a\ quasi\ ``up"\ quark}&:p_0=\sqrt{\bm{p}^2+m_{u_q}^2}\pm\mu_q,\\
{\rm a\ quasi\ ``down"\ quark}&:p_0=\sqrt{\bm{p}^2+m_{d_q}^2}\mp\mu_q,\\
\end{array}
\label{QuarkDR}
\end{equation}
where $u_q$ runs over $u$, $c$, $t$, while $d_q$ over $d$, $s$, $b$.
The extra potential terms $\mu_l$ and $\mu_q$ are given by 
\begin{equation}
\begin{array}{cc}
\mu_l=\displaystyle\frac{m_l}{2}, &\mu_q=\displaystyle\frac{\sqrt{m_{u_q}^2+m_{d_q}^2}}{2},
\end{array}
\label{DEFmu_q}
\end{equation}
where $m_l$ is the mass of charged lepton $l$, while $m_{u_q}$  and $m_{d_q}$ are the quark masses of the $q$-th generation with isospin $1/2$ and $-1/2$, respectively. 

We consider an effective field theory of these quasi fermions. 
The general form of Lagrangian density for a Dirac spinor $\psi_+$ generating dispersion relations (\ref{LeptonDR}) or (\ref{QuarkDR}) is given by
\begin{equation}
{\cal L}=\bar{\psi}_+[\gamma^\mu(i\partial_\mu-f_\mu)-m]\psi_+,
\label{DL}
\end{equation}
where 
\begin{equation}
f^\mu=f(1,\bm{0}).
\label{FPV}
\end{equation}  
A constant $f$ takes values $\pm\mu_l$ or $\pm\mu_q$, according as a quasi fermion is of lepton-type or of quark-type.
In a general Lorentz frame, $f^\mu$ takes the form
\begin{equation}
f^\mu=f\bar{u}^\mu=\frac{f}{\sqrt{1-w^2}}(1,-\bm{w}),
\label{CVP}
\end{equation}  
where $\bm{w}$ is a velocity vector of a new Lorentz frame with respect to the fiducial one.
The constant vector potential $f^\mu$ seems to break Lorentz invariance at the Lagrangian level. 
We also note that, compared to $a$-coefficients considered in the literature \cite{CK1,CK2}, $f^\mu$ is very large. 

Since the equation of motion of the quasi fermion satisfies the relation
\begin{equation}
(p-f)^2=m^2,
\label{ACEM2}
\end{equation}
we have two solutions  
\begin{equation}
p^0=\pm\sqrt{(\bm{p}-\bm{f})^2+m^2}+f^0.
\label{FDS}
\end{equation}
Ordinarily, the solution with a negative root square is not allowable since then the energy spectrum becomes unbounded below and the vacuum would be unstable. In this case, the hole interpretation applies to the negative root solution. Then we have the dispersion relation 
\begin{equation}
p^0=\sqrt{(\bm{p}\mp\bm{f})^2+m^2}\pm f^0,
\label{FAFDS}
\end{equation}
which satisfies
\begin{equation}
(p\mp f)^2=m^2,
\label{PDS2}
\end{equation} 
where the lower sign corresponds to a quasi-anti-fermion.
If the particle velocity is defined by the group velocity 
\begin{equation}
\bm{v}=\frac{\partial p^0}{\partial \bm{p}}
=\frac{\bm{p}\mp\bm{f}}{p^0\mp f^0},
\label{GVL}
\end{equation}
we see from (\ref{PDS2}) that $\bm{v}^2<1$ for a massive quasi-fermion and $\bm{v}^2=1$ for a massless quasi-fermion, in accordance with the ordinary relativity theory.
In terms of the particle velocity $\bm{v}$, the canonical 4-momentum $p^\mu$ is rewritten in the particle picture as
\begin{equation}
p^\mu=k^\mu\pm f^\mu,
\label{PDS}
\end{equation}
where
\begin{equation}
k^\mu=(\frac{m}{\sqrt{1-v^2}},\frac{m\bm{v}}{\sqrt{1-v^2}})
\label{KMV}
\end{equation}
is the kinematical 4-momentum.

One of the noticeable features of (\ref{PDS}) is the breakdown of the positivity of energy when $|m|<f^0$.
However, the vacuum will not fall into catastrophe, since negative energies are bounded below.

Another feature concerns the interrelation of potential terms $\pm f^\mu$ for a quasi-fermion and an quasi-anti-fermion.
We can interpret this term as CPT-violating, if $f^\mu$ is assumed invariant under CPT-conjugation. This observation is easily confirmed from the expression (\ref{DL}), since $\bar{\psi}_+\gamma^\mu\psi_+$ is CPT odd. 
In SME, this interpretation refers to the ``particle Lorentz transformation".
We will not make use of the Lorentz transformations of this type except for discussing the CPT-property of quasi-fermions.

As first noticed in SME \cite{CK1}, the quasi fermion operator $\psi_+$ and the ordinary Dirac operator $\psi$ satisfy the following relation 
\begin{equation}
\psi(x)=e^{if\cdot x}\psi_+(x).
\label{PsiEx}
\end{equation}
From the mode expansion
\begin{equation}
\begin{array}{rl}
\psi_+(x)=&\sum_{s\bm{p}}q_{s\bm{p}}U_{s\bm{p}}e^{-ipx}+\bar{q}^\dagger_{s\bm{p}}V_{s\bm{p}}e^{i\bar{p}x}\\
=&\sum_{s\bm{k}}a_{s\bm{k}}u_{s\bm{k}}e^{-i(k+f)x}+b^\dagger_{s\bm{k}}v_{s\bm{k}}e^{i(k-f)x},
\end{array}
\end{equation}
we see that the annihilation operators $q$ and $\bar{q}$ for quasi-fermions are expressible in terms of those $a$ and $b$ for ordinary fermions. This observation implies that the vacuum is common for both $\psi$  and $\psi_+$, and a quasi fermion with momentum $p^\mu$ is equivalent to an ordinary fermion with momentum $k^\mu=p^\mu-f^\mu$, while a quasi anti-fermion with momentum $p^\mu$ is equivalent to an ordinary anti-fermion with momentum $k^\mu=p^\mu+f^\mu$.

The scattering amplitudes are therefore calculable in terms of the ordinary fermion $\psi$ by replacing the initial and the final states with those corresponding to ordinary fermions: 
\begin{equation}
\langle p'_i |S_q|p_i\rangle=\langle k'_i|S| k_i\rangle,
\label{AmpTr}
\end{equation}
where the scattering operator $S$ for ordinary fermions is obtained from $S_q$ for quasi-fermions by the field redefinition (\ref{PsiEx}).
If $S$ is Lorentz invariant, 4-momentum conservation will give to an amplitude the factor
\begin{equation}
\delta^4\left(\sum_{i=1}^{N'}{p'}_i^\mu-\sum_{i=1}^{N}p_i^\mu-(F'-F)f^\mu\right),
\label{4MC}
\end{equation}
where  $F$ and $F'$ are the fermion numbers in the initial and the final states, respectively. Therefore, if interactions conserve the fermion number, both the canonical and the kinematical 4-momenta conserve. 

When there are various kinds of quasi fermions with potentials $f_\alpha^\mu$, the total canonical momentum conserves only if the condition 
\begin{equation}
\sum_\alpha (F'_\alpha- F_\alpha)f^\mu_\alpha=0
\label{QMC}
\end{equation}
holds, where $F_\alpha$ and $F'_\alpha$ are the numbers of quasi fermions of species $\alpha$ appearing in the initial and the final states.

Incidentally, if the origin of $f^\mu_\alpha$ is due to some spontaneous Lorentz violation as discussed in the preceding paper \cite{KN}, it is natural to suppose that all the space components of constant vectors $f^\mu_\alpha$ disappear simultaneously in the  fiducial Lorentz frame. 
Then the condition (\ref{QMC}) reduces to
\begin{equation}
\sum_\alpha (F'_\alpha- F_\alpha)f_\alpha=0.
\label{RQMC}
\end{equation}
We presently see that the interactions by charged weak currents do not satisfy (\ref{RQMC}) in SMR.

When we move to another Lorentz frame, the dispersion relations (\ref{LeptonDR}) and (\ref{QuarkDR}) are altered due to the transformation  of $f^\mu$. The choice of the fiducial Lorentz frame often gives rise to a problem in SME. 

In the Big Bang Cosmology there exists a special reference frame in which the cosmological background radiation looks isotropic. Though it may be natural to take this frame as fiducial, a rest frame of the Sun is customary taken in the literature \cite{KR}.

In the context of SMR, on the other hand, the local phase transformations for $\psi_{\pm\alpha}$ can calibrate any Lorentz frame into the fiducial one. Then no physical difference arises from the choice of the fiducial Lorentz frame.

\section{Standard-model redefinition}
In an effective field theory of the unified model of fermions, the quasi neutrinos $\nu_{l+}$ and the charged quasi leptons $l_-$ have the free Lagrangian
\begin{equation}
\begin{array}{rl}
{\cal L}_L=\displaystyle\sum_{l=e,\mu,\tau} &\bar{\nu}_{l+}\gamma^\mu(i\partial_\mu- f_{l\mu})\nu_{l+}\\
+&\bar{l}_-[\gamma^\mu(i\partial_\mu+ f_{l\mu})-m_l]l_-.
\label{FLL}
\end{array}
\end{equation}
Similarly, the quasi quark doublets in the isotropic representation are described by the effective free Lagrangian:
\begin{equation}
\begin{array}{rl}
{\cal L}_Q=\displaystyle\sum_{q=1}^3 &\bar{u}_{q+}[\gamma^\mu(i\partial_\mu- f_{q\mu})-m_{u_q}]u_{q+}\\
+&\bar{d}_{q-}[\gamma^\mu(i\partial_\mu+ f_{q\mu})-m_{d_q}]d_{q-},
\label{FQL}
\end{array}
\end{equation}
where we have assumed that $ f_{q\mu}$ are independent of the color degrees of freedom which have been already suppressed.

If these $f$-potentials are considered in the context of SME, the conservation of energy seems to disagree with the conventional form in some reactions. 
For example, we consider the decay of a $\mu$-on at rest,
\begin{equation}
\mu^-\rightarrow \nu_\mu+e^-+\bar{\nu}_e.
\label{MuDecay}
\end{equation}
According to the dispersion relations (\ref{LeptonDR}), the energy conservation seems to demand that
\begin{equation}
m_\mu/2=(|\bm{p}_1|+m_\mu/2)+(\omega_e-m_e/2)+(|\bm{p}_2|-m_e/2),
\label{ECMuDecay}
\end{equation}
where $\bm{p}_1$ and $\bm{p}_2$ are momenta of $\nu_\mu$ and $\bar{\nu}_e$, respectively.
The energy of $\mu$-on at rest is $m_\mu/2$ in view of (\ref{LeptonDR}).
Since the right-hand side of (\ref{ECMuDecay}) is greater than $m_\mu/2$, the $\mu$-on decay appears impossible, otherwise the energy conservation would be violated.

This conclusion stems from an assumption that the interaction terms for charged weak currents are the same as those given in the standard model.
Therefore, if we expect that the unified model of fermions effectively generates the standard model, an effective interaction term for the charged weak current should have the form
\begin{equation}
\frac{g}{2\sqrt{2}}e^{2if_l\cdot x}\bar{l}_-\gamma^\mu(1-\gamma_5)\nu_{l+}W_\mu.
\label{EWIL}
\end{equation}
We note that, among the various interaction currents appearing in the standard model, the maintenance of the conventional energy-momentum conservation requires extra phase factors only for the charged weak currents.

The form of interaction Lagrangian (\ref{EWIL}) indicates the breakdown of the translational invariance and therefore of the energy-momentum conservation in the quasi-fermion picture. Even in this case, however, the free Lagrangian (\ref{FLL}) and the modified charged weak current (\ref{EWIL}) recover their ordinary forms by the local phase transformations
\begin{equation}
\begin{array}{cc}
\nu_l=e^{+if_l\cdot x}\nu_{l+},&l=e^{-if_l\cdot x}l_-,
\end{array}
\label{PTL}
\end{equation}
which shows that the theory still satisfies Poincar\'{e} invariance in the ordinary picture of leptons.

The above observation suggests that the effective field theory for the unified model of fermions should have a form which is obtained from the standard model by some local linear phase transformations for spinor fields. We have called such a form of the standard-model the standard model redefinition (SMR).
In SMR, only the kinematical 4-momentum  conserves. 

The canonical 4-momentum and the kinematical 4-momentum are definable also by the conserved energy-momentum tensors for free quasi-fermions
\begin{equation}
T_C^{\mu\nu}=\sum_\alpha\bar{\psi}_{\pm\alpha}\gamma^\mu i\partial^\nu\psi_{\pm\alpha}, 
\label{CEMT}
\end{equation} 
and
\begin{equation}
T_K^{\mu\nu}=\sum_\alpha\bar{\psi}_{\pm\alpha}\gamma^\mu i\nabla_{\pm\alpha}^\nu\psi_{\pm\alpha},
\label{PTEMT}
\end{equation} 
respectively, where $\nabla_{\pm\alpha\mu}=\partial_\mu\pm if_{\alpha\mu}$.
According to (\ref{CEMT}) or (\ref{PTEMT}), a free quasi-fermion has the expectation value
\begin{equation}
\langle \int d^3xT_C^{\mu\nu}\rangle_\pm=\frac{k^\mu (k\pm f)^\nu}{\omega},
\label{DVEV}
\end{equation} 
or
\begin{equation}
\langle \int d^3xT_K^{\mu\nu}\rangle_\pm=\frac{k^\mu k^\nu}{\omega}.
\label{QFVEV}
\end{equation} 
When the charged weak current interactions are switched on and off adiabatically, the canonical energy-momentum tensor (\ref{CEMT}) does not conserve in SMR, since interaction terms have explicit space-time dependence as exemplified by (\ref{EWIL}).

On the other hand, the kinematical energy-momentum tensor (\ref{PTEMT}) still conserves, since the local phase transformations which remove the extra phase factors of interaction currents in SMR reduce (\ref{PTEMT}) to the ordinary form of canonical energy-momentum tensor. 

\section{Effects on gravity and electromagnetism}
Since the energy-momentum tensor is also the source of gravity, the $f$-potentials may affect gravity. 
Due to the Bianchi identity, however, the source term of the Einstein equation should be the conserved symmetric tensor, which is given by 
\begin{equation}
T^{\mu\nu}=\frac{1}{4}\sum_\alpha\bar{\psi}_{\pm\alpha}(\gamma^\mu i\nabla_{\pm\alpha}^\nu+\gamma^\nu i\nabla_{\pm\alpha}^\mu)\psi_{\pm\alpha}+{\rm h.c.}
\label{CQLEMT}
\end{equation}  
in a flat space-time. 
This expression also coincides with that obtained from the functional derivative of the free quasi-fermion Lagrangian in curved space-time with respect to the tetrad \cite{SW}, and gives the kinematical 4-momentum for a quasi fermion. 
Therefore we conclude that the $f$-potentials will not contribute to gravity, at least as the primary source.  

There is an argument in SME that the symmetric energy-momentum tensor does not conserve for a fermion with an $a$-coefficient \cite{CK1}. However, the conserved symmetric $T^{\mu\nu}$ is still definable by (\ref{CQLEMT}). The moral of this observation is to distinguish between the 4-momentum appearing in dispersion relations and that obtained from the conserved $T^{\mu\nu}$. 
The 4-momentum measured by experiments are based on the conservation laws, and therefore corresponds to the value of the conserved $T^{\mu\nu}$. On the other hand, the 4-momentum appearing in the dispersion relation or in the phase factor of a wave function is a canonical quantity, not a kinematical one.
Whether the quasi fermions may reveal the effect from Lorentz or  CPT violation depends on whether the phenomenon originates due essentially to the canonical phase of a quasi-fermion. Even for a quasi-fermion, the observational 4-momentum corresponds to the kinematical one in SMR. 

There is also an argument in SME that a gauge transformation in QED can not eliminate an $a$-coefficient. 
This statement seems apt to lead misunderstandings, since the explanation in the literature \cite{CK1} is not plain. We show that this conclusion is specific in SME, not applicable to SMR. 

There are two kinds of gauge transformations known in QED. One is that for spinors and the other is that for the electromagnetic potential. The former is called the gauge transformation of the first kind, while the latter the second kind \cite{Schiff}.

We consider in SME the following Lagrangian for the electron
\begin{equation}
{\cal L}_e=\bar{\psi_e}[\gamma^\mu(iD_\mu- a_\mu)-m_e]\psi_e-\frac{1}{4}F^2,
\label{QEDL}
\end{equation}
where $D_\mu=\partial_\mu-ieA_\mu$.
An $a$-coefficient is eliminable either by a gauge transformation of the first kind:
\begin{equation}
\psi_e=e^{-ia\cdot x}\psi_e^0
\label{GTF}
\end{equation}
or by a gauge transformation of the second kind:
\begin{equation}
A_\mu=A_\mu^0+a_\mu/e.
\label{GTS}
\end{equation}
Suppose that an $a$-coefficient is eliminated by a gauge transformation of the first kind. 
Then this transformation brings a local phase factor into an interaction term of the charged weak current:
\begin{equation}
\frac{g}{2\sqrt{2}}\bar{\psi_e}\gamma^\mu(1-\gamma_5)\psi_\nu W_\mu.
\label{CWC}
\end{equation}
The invariance of (\ref{CWC}) requires that the $W$-boson field should be transformed as 
\begin{equation}
W_\mu=e^{ia\cdot x}W_\mu^0.
\label{LFTforW}
\end{equation}
The second transformation in turn modifies the kinetic plus electromagnetic interaction terms of the $W$-boson: 
\begin{equation}
{\cal L}_W=-\frac{1}{2}W^{\mu\nu}{}^\dagger W_{\mu\nu}
+m_W^2W^\mu{}^\dagger W_\mu,
\label{WL}
\end{equation}
where $W_{\mu\nu}=D_\mu W_\nu-D_\nu W_\mu$. 
The invariance of (\ref{WL}) requires that the electromagnetic potential should be transformed as
\begin{equation}
A_\mu=A_\mu^0-a_\mu/e.
\label{GT2}
\end{equation}
The last gauge transformation of the second kind brings the $a$-coefficient once eliminated back into (\ref{QEDL}) again.
The conclusion is indeed the same when an $a$-coefficient is eliminated first by a gauge transformation of the second kind.
 
On the other hand, in the context of SMR, a charged weak interaction Lagrangian is given by (\ref{EWIL}), not by (\ref{CWC}).
In this case, an extra phase factor is removable simply by some local phase transformations for the quasi-electron and the quasi-neutrino, without affecting the $W$-boson sector.

\section{Neutrino oscillations}
Though all $f$-potentials are removable by the local phase transformations in SMR, since a field redefinition introduces space-time dependence into the particle identification, if fermion mixings occur, such as quark mixings or neutrino mixings, it  may not be obvious that the space-time dependence of particle identification does not induce physical effects. 
The result depends on whether the mixing is considered in the context of SME or SMR.

We suppose that neutrinos appearing in charged weak currents are not the quasi-neutrinos in 4-momentum eigenstates but linear combinations of them: 
\begin{equation}
\nu'_{l}=\sum_{l'}U_{ll'}\nu_{l'}.
\label{NM}
\end{equation}
Then SMR substitutes for $\nu'_l$ 
\begin{equation}
\sum_{l'}U_{ll'}e^{if_{l'}\cdot x}\nu_{l'+},
\label{NMTLV}
\end{equation}
where $U_{ll'}$ are elements of some unitary matrix.
Since we have
\begin{equation}
|\nu'_{l+}(x)\rangle=\sum_{l'}U_{ll'}e^{-i(p-f)_{l'}\cdot x}|\nu_{l'+}(0)\rangle,
\label{SVforN}
\end{equation}
the neutrino oscillation for two flavor mixing $\nu_\mu\leftrightarrow\nu_e$ is expressed by
\begin{equation}
|\langle\nu'_{\mu+}(0)|\nu'_{\mu+}(x)\rangle|^2=1-\sin^22\theta_{12}\sin^2\frac{\Delta k\cdot x}{2},
\label{NOsc}
\end{equation}
where $\theta_{12}$ is a mixing angle and  $\Delta k^\mu$ is the difference of two kinematical momenta of quasi-neutrinos in the detector frame: $\Delta k^\mu=k^\mu_1-k^\mu_2$.   
If quasi-neutrinos in 4-momentum eigenstates do not have different kinematical masses, 
then $\Delta k^\mu=0$, which implies that the $f$-potentials do not cause neutrino oscillations. 

On the other hand, if we consider in the context of SME, the phase factors in (\ref{NMTLV}) do not appear and neutrinos do oscillate.
In this case, however, the $f$-potentials introduce too large oscillations   and also do not have the energy dependence as observed in experiments nor reproduce $a$-coefficients assumed in the explanations of neutrino oscillations by Lorentz violation \cite{KM1,KM2,KM3,DK,DKM}.
These observations in turn also supports the view that an effective field theory of the unified model of fermions should be of TLV type.

In SME, there are arguments on the detection of Lorentz violation in the oscillations of neutral meson systems \cite{VAK1,VAK2,VAK3,KK}.
In the context of SMR, if the origin of CP-violation in neutral meson systems is supposed to be due to the quark flavor mixings characterized by the CKM matrix \cite{CKM}, the same result as neutrino oscillations will apply also to neutral meson systems. Then $f$-potentials will not directly contribute to CP or CPT violation for oscillations in neutral meson systems.

\section{Summary and conclusion}
Due to the nature of spontaneous symmetry breakdown, the largeness of $f$-potentials and the energy-momentum conservation in charged weak current processes, it is concluded that an effective field theory of the unified model of fermions should be of the standard-model redefinition (SMR) type, rather than of the standard-model extension (SME) type.

Then, the possibility to detect physical effects from the Lorentz- and CPT-violating $f$-potentials can be examined in the context of SMR.
We have considered this possibility from the aspects of gravitational effects and flavor-mixings.

Contrary to the first expectation, it is concluded that $f$-potentials will not cause any physical effects in the gravitational phenomena nor in the neutrino or neutral meson oscillations, which contrasts to the arguments in SME.
 
The notion of SMR in turn suggests a possible form of the effective interaction Lagrangian for the unified model of fermions.
Though it has not yet been investigated well, the form of a charged weak current (\ref{EWIL}) suggests that its derivation would be difficult, since extra phase factors do not allow perturbative treatments.

One question naturally arises of whether the result that the SLV of a local field theory generates an effective field theory of TLV type is specific only for a particular model. 

As a general argument, since SLV really breaks no symmetry at the operator level, it may be expected that a model of spontaneous Lorentz violation would show no measurable breakdown of Lorentz invariance.
This expectation is partly justified by the fact that the Lorentz-violating potentials in the unified theory of fermions have such a nature that they are removable by some phase transformations for effective fields. 

There remains a possibility that $f$-potentials may indirectly induce by interactions $a$-coefficients discussed in SME. The $a$-coefficients in SME are directly detectable by experiments and generally not eliminable without affecting the other interaction terms.
Furthermore, they are generally anisotropic and have simple forms only in a specific Lorentz frame. Therefore sidereal variations in  experimental observations can be the sign of Lorentz violation.
On the other hand, though $f$-potentials can be also anisotropic, they induce no physical effect and therefore no sidereal variation in experimental results.
In view of this qualitative difference, it will be difficult to imagine that the interactions induce from the $f$-potentials in SMR the $a$-coefficients in SME.

According to the observations made in this paper,
if the standard model is generated by spontaneous Lorentz violation of some underlying theory, the effective field theory can be of SMR type, rather than of SME type.
In particular, if the underlying theory is the unified model of fermions, 
it is highly probable that its effective field theory is of SMR type, due to the largeness of $f$-potentials. 
Then no Lorentz violation will be detectable in the unified theory of fermions, possibly except for the baryon asymmetry of the Universe, or  at least in the phenomena describable by an effective field theory.

\section*{Acknowledgements}
The author thanks H. Kunitomo for valuable comments and T. Kugo for useful correspondence.

\end{document}